\RequirePackage{fix-cm}
\documentclass{elsarticle}       % onecolumn (second format)
\usepackage{graphicx}
\usepackage{amsmath, amssymb}
\newtheorem{thm}{Theorem}

\newdefinition{rmk}{Remark}
\newproof{pf}{Proof}

\begin{document}

\title{Statistical Validity and Consistency of Big Data Analytics: A General Framework%\thanks{Grants or other notes
}


\author[BK]{B.~Karmakar}
\ead{bikramk@upenn.com}
\author[IM]{I.~Mukhopadhyay\corref{cor2}}
\ead{indranil@isical.ac.in}

\cortext[cor2]{Corresponding author}

\address[BK]{University of Pennsylvania, PA, USA}
\address[IM]{Indian Statistical Institute, Kolkata, India}

%\author{Bikram Karmakar \and \\ Indranil Mukhopadhyay %etc.}

%\institute{B. Karmakar \at
%              University of Pennsylvania, USA \\
%              \email{bikramk@upenn.com}           %  \\
%%             \emph{Present address:} of F. Author  %  if needed
%           I. Mukhopadhyay \at
%              Indian Statistical Institute, India\\
%   	\email{indranil@isical.ac.in}
%}


\begin{abstract}
Informatics and technological advancements have triggered generation of huge volume of data with varied complexity in its management and analysis.  Big Data analytics is the practice of revealing hidden aspects of such data and making inferences from it. Although storage, retrieval and management of Big Data seem possible through efficient algorithm and system development, concern about statistical consistency remains to be addressed in view of its specific characteristics. Since Big Data does not conform to standard analytics, we need proper modification of the existing statistical theory and tools. Here we propose, with illustrations, a general statistical  framework and an algorithmic principle for Big Data analytics that ensure statistical accuracy of the conclusions. The proposed framework has the potential to push forward advancement of Big Data analytics in the right direction. The partition-repetition approach proposed here is broad enough to encompass all practical data analytic problems.
\end{abstract}



\maketitle

\section{Introduction}
\label{intro}
`Big Data' presents itself with unique challenges in retrieving, storing and all the way to analysing the data.  Technological breakthrough makes generation and collection of huge volume of data possible in many fields like genetics, genomics, health care, customer service, informatics, to name a few. Among various challenges presented by the abundance of data, analysis of data is a well recognized hurdle. While the explosion of information allows us to know more about the process, appropriate methods or algorithms are essential to make correct inference or to reveal hidden pattern.

Recent advancements of technology and targeted methods to Big Data analytics give access to ample capacity for storing the data along with the skill of parallel computing. Much effort has been dedicated to extract information from Big Data in an efficient manner. From a practical standpoint, concern remains about the validity of results from analysis of Big Data. As attested by many recent articles, in most cases the inference based on such data is unacceptable and unreliable. For example, High dimension conventional classification methods are no better than random guesses \cite{Fan2008}. Understanding the output of Big Data analytics than to fixate on the technical aspect of it is the most important issue \cite{Fan2014,Fan2015,NewIdeas}, because the future decision making process depends only on this output. The aim of this article is to put forward a framework in order to establish  the acceptability of the learning from the Big Data. This framework also fits to the paradigm  of parallel computing and at the same time provides a robust statistical basis for practical application.

The classical statistical theory of data analysis has its roots in axioms of probability theory. With growing complexity of Big Data, statistical theory needs to be revisited  \cite{statBigData}, mainly due to violation of probabilistic independence or exchangeability situation. Statistics community has raised concerns about how the sound and carefully developed theory can help build a structure around it. In this article we exploit an algorithmic architecture used in practice to tackle Big Data and suggest an appropriate mathematical ground for analysis of such architecture.

We propose a partition and repetition approach in a general framework for statistical analysis of Big Data. This approach expands the horizon of standard statistical methods as well as opens new avenues for novel methods to encompass and tackle the challenges arisen due to the specific characteristics of Big Data. With the help of this general framework, we prove consistency and accuracy of the analytic results thus obtained. We have explained this theory through various examples that are usually required in common data analysis paradigm in respect of many fields. We hope that such a framework would help in further development of Big Data analytics.

\section{The divide and conquer algorithm}
Abundance of digital information is one way to explain what we today understand as `Big Data'. There are two aspects to the story. Firstly, human intuition suggests that accuracy of the answer to our question increases if we have more and more information. This intuition works backward; we start with a question, try to comprehend what data we might need to answer the question and then realize that relevant information exists somewhere in digitized format. The catch is that, this retrospective thought process assumes that the skill by which human intelligence finds this answer from the data is transferable to mechanical and algorithmic computing. Secondly, with huge volume of data we can find a question of interest from the data itself and then get the answer to the question. But the inherent complexity of available data makes this task difficult. This whole process is advertised as Big Data analytics.

Principle characteristics of Big Data are its volume, velocity, variety and complexity \cite{Characteristics}. All of them presents as unique challenges at a technical level of dealing with the data. At the hardware level we have reached a saturation point on the achievable clock pulse on a single processor. Rather, the growth in computing capacity is attained by increasing the number of threaded cores. Moreover, while storage capacity is fairly cheap and scalable, the RAM is not so. Recognizing this hardware restriction the state of the art algorithms (Hadoop, Amazon EC2) for Big Data analytics has adopted a partitioning based method.

However, in view of advancements in computing systems including storage and processing, need for new data analytic tools are required that are adaptive to new technologies \cite{petcu}. Building such statistical tools and algorithms for monitoring and analysis is needed to achieve success in Big Data analytics. Hence standard statistical methods should be revisited, modified, and validated in the light of scalability to extremely large scale data applications \cite{reed}.

Fisher et al.~\cite{BigDataInteracting} have identified the standard workflow of data analysis as, 1) acquiring data, 2) choosing an architecture, 3) shaping the data to the architecture, 4) writing and editing the code, and 5) reflecting and iterating on the results. The initial struggle is to adopt a suitable architecture for the data and map the collected data to that architecture. In this article, we are not focusing on this domain of analytics job. Rather the focus is on the later part of analysing the data. To address the problem of huge volume of data, the way is to partition it into small portions that are manageable by RAM, process the data in a parallel manner, and finally combine the processed information to produce the final output. This idea of partitioning has been used, although in a subtle way, in other areas of research, e.g., data mining \cite{buehrer, calders}, MCMC \cite{wang}. An extra benefit of this divide and conquer method is that such an algorithm easily adapts to the velocity of Big Data. Velocity contributes to new partitions which are to be analysed and then the inference is to be combined with the earlier output \cite{velocity}. Issues relating to variety and complexity are taken into account by the statistical methods and algorithms that are used in the analysis (see discussion of Section 4 and References therein).

\section{The framework}

\subsection{Sample space structure}

Classical theory of statistical analysis is a well developed area with sound theories. To establish a framework for Big Data analytics we naturally would like to fall back on those works. To begin with, we consider a sample space $(\mathcal{S}, \mathcal{A}_\mathcal{S})$ where $\mathcal{S}$ is the space of realized values of the data and $\mathcal{A}_\mathcal{S}$ is the sigma field associated with the sample space. We denote by $\mathcal{M}(\mathcal{S})$ the set of probability measures on $(\mathcal{S}, \mathcal{A}_\mathcal{S})$. Also let $\mathcal{M}^e(\mathcal{S}) \big(\subset\mathcal{M}(\mathcal{S})\big)$ be the set of probability measures with finite support. An observed data $X_{n \times d}$ can be identified by a probability measure $m_X$ on $(\mathcal{S}, \mathcal{A}_\mathcal{S})$, with a support having finite cardinality, defined as follows,
$$m_X(A) = \sum_{x_i \in A, i = 1}^n \frac{1}{n},$$
for any $A\subseteq \mathcal{A}_\mathcal{S}$ and $x_i\, (i=1,\ldots,n)$ is the $i$-th data point. To build a theory around it we would require a suitable metric on the space $\mathcal{M}(\mathcal{S})$. For example, if $(\mathcal{S}, \mathcal{A}_\mathcal{S})$ is a polish space then with Prokhorov metric ($d_\mathcal{M}$) we can put weak convergence on $\mathcal{M}(\mathcal{S})$. 

Till this point we have not considered any aspect of Big Data par se. Our aim is to build the ideology of Big Data analytics on this sample space structure. Identification of the realized data with an empirical measure on some sample space gives a broader ground to work on. In a Big Data set up, we hardly have any control on the generation of data. Thus unlike in classical statistical theory, where mostly we want to build better experimental designs to apply statistical methods, be it standard or novel, here we want to construct an algorithm that would work with the data generation process. This difference in approaches is subtle but central to how these two ideologies differ. 

\subsection{The problem approach}

Main goal of Big Data analytics is to extract information from the data, which is equivalent to getting information from an element in $\mathcal{M}^e(\mathcal{S})$. So we assume that a satisfactory data collection and mapping architecture exists. To develop a full framework, we introduce some definitions about functionality of data analysis. This is necessary to avoid the cumbersome details and technicalities of a particular scenario.

Extracted information of a data analysis can be viewed as an element in the result space ($\mathcal{R}$). A {\em problem approach} ($\rho$) is a function from $\mathcal{M}^e(\mathcal{S})$ to $\mathcal{R}$.  Based on this formulation of problem approach we can consider two classes of problem approaches as follows.

\paragraph*{Definition 1} {\em Inference Problem}: If the problem approach $\rho$ can be extended to a strictly larger subset of $\mathcal{M}(\mathcal{S})$ than $\mathcal{M}^e(\mathcal{S})$, then such a problem or problem approach is called an inference problem.

\paragraph{Definition 2} 
{\em Mining Problem}: If the problem approach $\rho$ can only be defined on $\mathcal{M}^e(\mathcal{S})$, then such a problem or problem approach is called a mining problem.\\

The usual examples of these two classes of problems are as follows. Parametric estimation and testing problems fall under the class of inference problems where the subset of $\mathcal{M}(\mathcal{S})$ under consideration is $\mathcal{M}^e(\mathcal{S})$ along with the parameter models. Clustering problem or outlier detection problem, on the other hand, are under the class of mining problems. In later sections, we shall discuss both these classes of problem approaches and their solutions in more details.

A technical assumption we need to have is that, one such problem approach is {\em viable} if the map 
\begin{equation}\label{eq:viableProblem}
\rho : \big(\rho^{-1}(\mathcal{R}), d_\mathcal{M}\big) \longrightarrow \big(\mathcal{R}, d_\mathcal{R}\big)
\end{equation}
is a continuous map, where $d_\mathcal{M}$ and $d_\mathcal{R}$ are appropriate metrics on respective spaces. A {\em viable problem approach ($\rho$)} then ensures that the problem is consistent in the number of samples as well as data points. This means that slight change in the data generation process ($\mathcal{M}(\mathcal{S})$) should not create substantial difference in the result ($\mathcal{R}$).

\subsection{Big Data Algorithm}

We now discuss various components of our proposed algorithmic structure of Big Data analytics. 

\subsubsection{Partitioning}
A naturally accepted strategy in analysing huge volume of data is to consider small parts of data at a time. Our formulation for Big Data analytics formulates this method of partitioning the data as a functional,
\begin{equation}
\begin{aligned}
&H_L : \mathcal{M}^e(\mathcal{S}) \longrightarrow \mathcal{M}^e(\mathcal{S})\times \cdots \times \mathcal{M}^e(\mathcal{S})\;\;\;\;{(L\; many)}\\
&H_L(m) = (m_1, m_2, \ldots, m_L),
\end{aligned}
\end{equation}

such that $(m_1, m_2, \ldots, m_L)$ is related to $m$ by,
\begin{equation}\label{eq:suppPartitioning}
\begin{aligned}
&supp(m) = \mathop{\cup}\limits_{i=1}^L supp(m_i);\\ &supp(m_i)\cap supp(m_j) = \emptyset,\;\;1\leq i\neq j\leq L.
\end{aligned}
\end{equation}
where $supp(m)$ denotes the support set of $m$ and $\emptyset$ denotes the empty set.

For convenience we write $supp(m_i) = \big(x_1^{(i)}, x_2^{(i)}, \ldots, x_{n_i}^{(i)}\big)$ for each $i$. For a fixed data $m$ (or $m\equiv X$) we would be given a problem approach $\rho$. Then the divide and conquer strategy would choose a partitioning functional $H_L$.

But to reduce the error in result due to partitioning, the strategy is to repeat $K (>1)$ times the partitioning; denote them by $H_{L,1}, H_{L,2},\ldots,H_{L,K}$. This type of algorithm we call as the {\em partition-repetition algorithm}. We now formulate this partition-repetition algorithm in a comfortable manner.

Let $\mathcal{H}_L$ be the set of all partitioning functionals $H_L$. A $\sigma$-field $\mathcal{A}_{\mathcal{H}_L}$ can be defined as the smallest $\sigma$-field on $\mathcal{H}_L$ such that the functions $f_{i,j}(\cdot)$ on $\big(\mathcal{H}_L, \mathcal{A}_{\mathcal{H}_L}\big)$ to $(\mathcal{S}, \mathcal{A}_\mathcal{S})$ are measurable for any choice of $m \in \mathcal{M}^e(\mathcal{S})$, where
$$f_{i,j}\big(H_L(m)\big) = x_j^{(i)}\;\;j = 1, 2, \ldots, n_i;\;i = 1, 2, \ldots, L.$$
Then the strategy of analysing data of unmanageable size by partition-repetition algorithm can be understood as a probability measure $P_{H_L}$ on the measurable space $\big(\mathcal{H}_L, \mathcal{A}_{\mathcal{H}_L}\big)$. More precisely  $\{H_{L,1}, H_{L,2},\ldots,H_{L,K}\}$ would be viewed as a random sample from the probability measure space $\big(\mathcal{H}_L, \mathcal{A}_{\mathcal{H}_L}, P_{H_L}\big)$. For simplicity of notation let us denote by $\rho^L$ the map,
$$
\rho^L:(m_1, \ldots, m_L) \longmapsto (\rho(m_1),  \ldots, \rho(m_L)) \;\;{for}\;m_i \in \mathcal{M}^e(\mathcal{S});\;
$$
for $i=1, 2, \ldots, L$.
Then a single random sample $H_L$ from the probability distribution $P_{H_L}$ provides us $L$ results $\rho^L\big(H_L(m)\big)$, which are $L$ elements from $\mathcal{R}$. With a random sample $H_{L,1}, H_{L,2}, \ldots, H_{L,K}$ from the distribution, the set of results we get using the problem approach $\rho$ is 
\begin{align*}
\{{R}_{k,l}^*\}_{k = 1, 2, \ldots, K;\; l = 1, 2, \ldots, L}  & = \big\{{R}_{k,1}^*, {R}_{k,2}^*, \ldots, {R}_{k,L}^*\big\}_{k=1,2,\ldots,K}\\
& = \big\{\rho^L\big(H_{L,k}(m)\big)\big\}_{k=1,2,\ldots,K}.
\end{align*}

This framework also encompasses the case where rather than partitioning one chooses to sub-sample. In that case we would get rid of the extra restriction in equation \eqref{eq:suppPartitioning} on the functional $H_L$. Popular algorithms of Bootstrap and Bag-of-Little-Bootstraps (BLB) \cite{kleiner2014} are covered in this framework.

\subsubsection{Combining}
Next critical part of the algorithm is combining the results obtained above, $\big\{{R}_{k,l}^*\big\}_{k = 1, 2, \ldots, K;\; l = 1, 2, \ldots, L}$ in order to arrive at a final result. Let $C_{KL}$ be the combining map that takes all the results from the collection and gives the final result. The triplet $\big(\rho, P_{H_L}, C_{KL}\big)$ can be called a {\em solution} to a Big Data problem.

Now it remains to understand the viability of the solution. We have put a stable condition of continuity in equation \eqref{eq:viableProblem} on problem approach $\rho$ as a {\em viable problem approach}. Proper behaviour of the pair $\big(P_{H_L}, C_{KL}\big)$ would ensure an accurate solution to the problem $\rho$ for $m$.

We focus on the case where $C_{KL} := C^2_K \circ C^1_L$ works in two stages. In the first stage $C^1_L$ works on each partition ($k$) to collect the results 
$${R}_k^* := C^1_L\big(\{{R}_{kl}^*\}_{l=1, 2, \ldots, L}\big)\;\;\;{for}\;k=1, 2, \cdots, K.$$
This $K$-tuple is combined by $C^2_K$. For a fixed data $m$ when $C^1_L$ is a measurable map, the randomness of $\{H_{L,1}, H_{L,2},\ldots, H_{L,K}\}$ makes the collection $\{{R}_1^*, {R}_2^*, \ldots, {R}_K^*\}$ an independently and identically distributed (i.i.d.) sample on the measure space $\big(\mathcal{R}, \mathcal{A}_{\mathcal{R}}\big)$. This formulation of the solution $\big(\rho, P_{H_L},  C^2_K \circ C^1_L\big)$ provides an opportunity to use rich statistical theory in data analytics.

In the general case, the result space can be quite complicated (we shall give concrete examples in later section). Rather than dealing with the space $\mathcal{R}$ itself it would be better to work with real numbers. This is achieved by an evaluation function $ev: \mathcal{R} \longrightarrow  \mathbb{R}^N$ for some fixed $N$ belonging to the set of natural integers. Then, viability of the choice of $P_{H_L}$  can be understood  using the evaluation function of the result space $\mathcal{R}$. For a given data $m$ and a problem approach $\rho$, we call a partitioning probability measure $P_{H_L}$ to be {\it viable} under the first stage combining operator $C^1_L$ if,
\begin{equation}\label{eq:via2}
\int ev \circ C^1_L\big(\rho^L(H_L(m))\big) dP_{H_L} = ev \circ \rho(m).
\end{equation}
This condition means that the probability measure $P_{H_L}$ and the combining method $C^1_L$ are compatible with each other for the problem $\rho$. If we do infinitely many repetitions of our partition-repetition based algorithm, the combining method $C^1_L$ will give equivalent performance as the one we would have got if we could apply $\rho$ on the data $m$.

The second stage of combining method $C^2_K$ operates on the collection of first stage result by combining ${R}_1^*, {R}_2^*, \ldots, {R}_K^*$ to get the solution 
$${R}_K^{**} := C^2_K\big(\{{R}_k^*\}_{k = 1, 2, \ldots, K} \big).$$
Now the viability of $C^2_K$ is based on the comparison of  ${R}_K^{**}$ with $\rho(m) = {R}^*$ (say). Here we present the soundness of the algorithm of partitioning and combining through the following theorem.

\begin{thm}\label{thm:vol}
For a Big Data solution $\big(\rho, P_{H_L},  C^2_K \circ C^1_L\big)$, if $P_{H_L}$ is a viable partitioning method under combining method $C^1_L$ (i.e., equation \eqref{eq:via2} is satisfied) and convergence in $ev$ is equivalent to that of in $\mathcal{R}$, then
there exists a second stage combining method $C^2_K$, such that ${R}_K^{**} \longrightarrow {R}^*$ almost surely in $P_{H_L}$.
\end{thm}
%\paragraph{Proof.}
\begin{pf}
Define $C^2_K$ on $\mathcal{R}\times \mathcal{R} \times \cdots \times \mathcal{R}$ ($K$ times) as follows,
$$C^2_K(R_1, R_2, \ldots, R_ K) := \arg \min_{\{R_k\}_{k = 1, 2, \ldots, K}} \big|\big| ev\circ R_i - ev \circ {R}^*\big|\big|.$$
Let us use the notations $Y_k = ev \circ {R}_k^*$, $Z_K =  ev \circ {R}_K^{**}$ and $\mu = ev \circ {R}^*$. Since $\{{R}_k^*\}_{k\geq 1}$ is an i.i.d. sample, by strong law of large numbers as equation \eqref{eq:via2} holds, for all $\epsilon > 0$ with $P_\mathcal{R} := P_{H_L}\circ C^{1^{-1}}_L \circ ev^{-1}$,
$$P_\mathcal{R}\Big(\cup_{k_0 = 1}^\infty \cap _{K \geq k_0} \Big(\Big|\Big| \frac{1}{K}\sum_{k=1}^K Y_k - \mu\Big|\Big| < \epsilon\Big)\Big) = 0.$$
Now using the fact that $|| \sum_{k=1}^K Y_k/K  - \mu|| \geq ||Z_K - \mu||$ and definition of $C_K^2$, the above holds with $\sum_{k=1}^K Y_k/K $ replaced by $Z_K$. Since convergence in $(\mathcal{R}, d_\mathcal{R})$ is equivalent to that in $\big(ev \circ \mathcal{R}, ||\cdot||\big)$, rest of the argument follows as by assumption convergence in $(\mathcal{R}, d_\mathcal{R})$ is equivalent to that in $\big(ev \circ \mathcal{R}, ||\cdot||\big)$.
\end{pf}

The theorem above deals with the volume aspect of Big Data. It says that even if the data is unmanageable to be processed practically, we can adopt partition-repetition approach to get a good solution. It is also not passed our attention that the number of combination rules may be more than two, but the final convergence of results requires some more assumptions and strong theorems in the dependence set up.

Next we also need to answer the question which is more of classical statistical in nature. If the velocity of the data provides us more and more information of specific form, is the partition-repetition algorithm able to extract that information? The following theorem tells us if that is the case, we would be able to choose a partitioning measure and a sequence of combining methods that gives the final result.

\begin{thm}\label{thm:velocity}
Let $\{m_n\}_{n\geq 1} \in \mathcal{M}^e(\mathcal{S})$ and $m \in domain \; of \; \rho$. Suppose the problem approach $\rho$ is viable on its domain and $m_n \longrightarrow m$. If conditions of Theorem \ref{thm:vol} hold for the sequence of solutions $\big(\rho, P_{H_L,n},  C^2_{K,n} \circ C^1_{L,n}\big)$, then there exists a sequence of integers $\{k_n : n \geq 1\}$ and a $P_{H_L}$ such that, for $n\geq 1$, $P_{H_L,n}$ is absolutely continuous with respect to $P_{H_L}$ with 
$$\Big|\Big| ev \circ C^2_{k_n,n}\circ C^1_{L,n}\big\{\rho^L(H_{L,k}(m_n)\big)\big\}_{k=1,2,\ldots,K}  - ev \circ \rho(m)\Big|\Big| \longrightarrow 0,$$
 as $n\rightarrow \infty$ almost surely in $P_{\mathcal{R}}$.
 \end{thm}
% and
%$$\big|\big|ev\circ \hat{\hat{R}}_{k_n,n} - ev \circ \rho(m)\big|\big| <\epsilon \;\;almost\;surely\;\;\big[P_{H_L}\big]$$
%where $\hat{\hat{R}}_{K ,n} = C^2_{K,n}\circ C^1_{L,n}\Big(\big\{\rho^L\big(H_{L,l}(m_n)\big)\big\}_{k=1,2,\cdots,K} \Big)$.

%\paragraph{Proof.} 
\begin{pf}
Define $P_{H_L}(\cdot) = \sum_{n=1}^\infty  P_{H_L,n}(\cdot)/2^n$. Let us denote,
$${R}_{K ,n}^{**} = C^2_{K,n}\circ C^1_{L,n}\Big(\big\{\rho^L\big(H_{L,k}(m_n)\big)\big\}_{k=1,2,\ldots,K} \Big).$$
Then for every $\epsilon (>0)$, by Theorem \ref{thm:vol} and equation \eqref{eq:viableProblem} there exists a sequence $\{k_n(\epsilon):n \geq 1\}$ and $N\geq 1$  such that for all $n\geq N$,
$$\big|\big| ev \circ {R}_{k_n(\epsilon),n}^{**} - ev \circ \rho(m)\big|\big| < \frac{\epsilon}{2^n},$$
almost surely in $P_{\mathcal{R},n} = P_{H_L,n}\circ C^{1^{-1}}_{L,n} \circ ev^{-1}$. %\;\;{almost\; surely\; in}\;\big[P_{\mathcal{R},n} = P_{H_L,n}\circ C^{1^{-1}}_{L,n} \circ ev^{-1}\big].$$
Choosing $\epsilon$ as rationals, result follows from Cantor's diagonal argument.
\end{pf}

Both these results are of existential nature rather than being instructive for practice. Although little abstract in their formulation, these theorems form the basis of the methods that would be applied in practice. Study on combining methods is not new to statistics. This framework enforces the importance of various combining methods along with partitioning methods in the light of Big Data analytics.

The power of this kind of theory is that we do not put any hard and fast regularity condition on the data or the data generation process. Theorem \ref{thm:velocity} only requires that the data collected eventually amounts to some specific information.

\section{Illustrative Examples}
An analyst's job and a statistician's work differ in a crucial way. An analyst is more concerned with how to extract information from the data available. This work is referred to as number crunching. A statistician is concerned about the quality of the extracted information sometimes taking for granted the effort of extracting the information. In a Big Data scenario where importance of analyst's job comes more into the limelight, a statistician could provide support by accepting some compromise on their ideology. In this section we illustrate the formulation developed above through some standard data analytic problems.

We first consider a few problems where the solution $\rho(m_n)$ can be calculated without any error from partitioning based algorithm. Here we specify by subscript $n$ the size of the data. In these examples it is enough to consider $P_{H_L}$ to be some degenerate probability distribution of convenience and we only require a single sample ($K=1$) from it.

\paragraph{Calculating sample mean}
Here $P_{H_L}$ can be any distribution that partitions the data into manageable balanced pieces. Then for $\rho(m_n) := (\int x\, dm_n, n)$ the combining method shall be,
$$C_L^1(\{(\bar{x}_i, n_i)\}_{i=1, 2, \ldots, L}) = \Big(\frac{\sum_i n_i \bar{x}_i}{\sum_i n_i}, \sum_i n_i\Big).$$
A little tweak in these definitions allows us to calculate many other descriptive statistics like weighted means, dispersion measures and also some robust measures for central tendency.

\paragraph{Sorting}
 To get a Big Data solution to the sorting problem we can define a partitioning $P_{H_L}$ as a degenerate distribution such that it divides the data $m_n$ into $L$ parts based on a sequence $bound_0 < bound_1 <\cdots < bound_L$ as,
$$bound_{i-1} \leq \{x_j^{(i)}\} < bound_i\;\;for \;i=1, 2, \cdots, L.$$
The choice of the sequence $\{bound_i\}$ should be such that the individual parts are of manageable sizes. With $\rho$ providing us with a sorted array, the combining stage should simply concatenate the ordered parts, i.e.,
$$C_L^1 (\{{R}_l^*\}_{l=1, 2, \ldots, L}) := ({R}_1^*, {R}_2^*, \cdots, {R}_L^*).$$

Similar solutions of the above type are obvious for problems like searching, calculating extreme statistics ($x_{(1)}, x_{(n)})$, constructing a histogram  etc. Most of the time these simple problems are only intermediate steps towards more challenging problems of data  analytics.
 
Some solutions to more standard problems of Big Data analytics are discussed in brief below. First few examples are inference problems while the later ones are mining problems. We assume that the data are cleaned and dressed for the purpose at hand. We avoid discussing the technical aspects of implementing these algorithms in practice, though in a few examples we shall provide references to available literature that has more focus on detailed analysis of the algorithms.

\paragraph{Estimation}
The problems of modeling (nonparametric, parametric, time series or even Bayesian) come under the radar of inference problem. Based on the requirements of the solution (e.g., unbiasedness, minimum variance, consistency) there would be different Big Data solutions to the problem approach $\rho$. Many of the times it suffices to consider $P_{H_L}$ as a random partitioning measure of the data, although while considering spatial and/or temporal data more clever partitioning measure would be required to satisfy viability condition like equation \eqref{eq:via2}.

Let us consider the problem of finding maximum likelihood estimate (MLE) for a parameter based on some algorithm (say, EM algorithm or Newton-Raphson or Fisher's Scoring etc.). The scenario is that, we have a statistical model in mind where the number of parameters is fixed. Then partitioning the data simply breaks the objective function (log-likelihood function) into $L$ parts. Consequently an intuitive choice of the combining method $C_{KL}$ would be whichever of the results from partitions maximizes the whole objective function. Although this method does not ensure the MLE for the data, but in practice we are hardly concerned about theoretical properties like efficiency; the estimate found by this method is acceptable.

\paragraph{Testing}
Consider a test function $\rho$ that provides p-value for testing $H_0$ against $H_1$. Then based on random partitioning of the data into balanced parts, a conservative combining algorithm \cite{combinePvals} for the corresponding solution can be
$${R}_k^* := C_L^1\big(\{{R}_{lk}^*\}_{l = 1, 2, \ldots, L}\big) = \min_{l=1,2, \ldots, L} {R}_{lk}^*, \text{for}\;k = 1, 2, \ldots, K,$$
and
$${R}^{**} := C_K^2\big(\{{R}_k^*\}_{k = 1, 2, \ldots, K}\big) = median \{{R}_k^*\}_{k = 1, 2, \ldots, K}.$$

\paragraph{Variable Selection}
The context in which variable selection problem has been  addressed in recent literature is sometimes too idealistic for Big Data paradigm, although there are some promising methods. The data generation process is assumed to provide information on a set of response variables and a fixed set of regressors. We might be interested in a subset of these variables which have effect on the responses. The quality of the selected variables can be assessed by proportions of the variables wrongly selected. In a situation where assumption of homoscedastic uncorrelated linear model is valid, Barbar and Candes \cite{knockoff} proposed a method to select variables with a control on the proportion of falsely discovered variables. This method is no doubt computationally heavy. The partition-repetition philosophy can be used to adapt this algorithm to achieve the same goal in current context.\\

If the data generation process is well controlled, the above inference problems and solutions make sense. Some recent works are available in the area of regression \cite{Battey,splitConquer} focusing on divide and conquer methods. Unfortunately spurious correlations, noisy data etc.~are very common in Big Data perspective. In that case these naive solutions can be hugely mis-representative of the actual truth.  Data mining problems are more relevant in such a scenario. In a mining problem we are interested in the data itself without having to make any modeling assumption. Possible Big Data solutions to a few mining problems are discussed below.

\paragraph{Clustering}
An elaborate and critical discussion on clustering problem in view of Big Data analytics can be found in recent article by the authors  \cite{combineCluster}. In brief, the combing method would identify the unique clusters from the set $\{{R}_{lk}^*\}_{l=1,2,\ldots,L}$ based on a decision function that tells us to combine two results when they seem to form a single data cloud. The second stage is to make stable clusters based on some measure from the $K$ sets of clusterings $\{{R}_k^*\}_{k = 1, 2, \ldots, K}$.

\paragraph{Outliers Detection}
Based on a random partitioning measure $P_{H_L}$ and a problem approach $\rho$ that separates the outliers ($m_n^o$) and the data  ($m_n^d$) section, (i.e., $\rho(m_n) := (m_n^d, m_n^o)$), the combining method $C_L^1$ would check the structure of the outliers from the individual parts and get the outliers from the whole part. The method should check if outliers from one part belongs to the data section of some other part and also if outliers from all the parts together form some data section. Second stage of combining would then pick out the stable outliers from all repetitions.

Ramaswamy et al.~\cite{combineOutlier} discuss another Big Data solution to this mining problem based on a different partitioning method based on clustering the data and van Stein et al.~\cite{vanStein} propose local subspace-based solution to outlier detection problem, which applies a combining strategy using global neighbourhoods. These methods can be viewed as special cases of our proposed framework.

\paragraph{Classification}
First we consider the $k$-NN classifier, where $\rho$ finds the $k$ nearest neighbours of a test data point ($x$) as,
$$\rho(m_n) := ((x_{(i)}, d(x,x_{(i)}))_{i=1,2,\ldots, k})\;\;$$
$$\begin{aligned}s.t.\;\;d(x,x_{(1)}) &\leq \cdots \leq d(x,x_{(k)})\\
&\leq min \{d(x,x_i); x_i \in X\setminus \{x_{(1)}, x_{(2)},\ldots, x_{(k)}\}\}.\end{aligned}$$ Based on any partitioning $P_{H_L}$, then the problem is exactly solvable in a single repetition with a combining operator that picks the $k$ data points nearest to $x$ among the $L\times k$ points. Subsequently the classifier is contracted on a second algorithm that simply checks for the maximum number of representatives in these $k$ data points from each of the classes.

Another celebrated class of classifiers is decision trees. A relevant combining operator of decision trees based on partition of the data is proposed by Hall et al.~\cite{combineDecision}.

\section{Discussion}

Data is the lubricant that drives the machinery of statistics. It is no longer a topic of debate that the way data is generated and collected in modern times is drastically different from what statisticians are used to deal with. Statistics should adapt to this change and thereby assist the masses of data analytic work.

The main contribution of this article is suggesting a basis of statistical theory for present day data analytic works. In composing the theory we have tried to stay true to the practical nature of a data science job. This formulation proposes a divide and conquer algorithm (either partition-repetition or subsampling method). More importantly it respects the fact that more often than not we have no control on the data generation process. We have also tried to encompass all possible data analytic problems. A range of such data analytic problems are discussed in perspective of our formulation. 

Successful use of statistical theory in data analysis would require understanding the field of `Big Data'. Rather than being insistent on developing methods and elaborate theories based on idealistic assumptions, we have kept their applicability in mind. Our proposed framework encompasses statistical analyses of majority of problems in view of complex characteristics of Big Data and can be extended further keeping its compatibility with modern advances in computational world.

\end{document}